\documentclass[aps,prl,twocolumn,longbibliography,nopacs,floats,nofootinbib,superscriptaddress]{revtex4-2}
\usepackage[english]{babel}
\babeladjust{autoload.bcp47 = on}
\renewcommand{\selectlanguage}[1]{}
\usepackage{graphicx}
\usepackage{dcolumn}
\usepackage{bm}
\usepackage{amsmath}
\usepackage{color}
\usepackage{amssymb}
\usepackage{ulem}
\usepackage{xcolor}
\usepackage{graphicx} 
\usepackage{braket}

\definecolor{cardinal}{rgb}{0.6,0,0}
\definecolor{darkgreen}{rgb}{0,0.4,0}
\definecolor{golden}{rgb}{0.92, 0.7, 0}
\definecolor{midnight}{rgb}{0, 0, 0.5}
\definecolor{darkblue}{rgb}{0, 0, 0.7}

\def\he4{$^4$He}
\def\hel3{$^3$He}
\def\Am3{\AA$^{-3}$}
\def\beq{\begin{equation}}
\def\eeq{\end{equation}}

\newcommand{\xh}{{\hat x}}

\newcommand{\zh}{{\hat{\bf z}}}

\newcommand{\cd}{c^{\dag}}
\newcommand{\be}{\begin{equation}}
\newcommand{\ee}{\end{equation}}
\newcommand{\bea}{\begin{eqnarray}}
\newcommand{\eea}{\end{eqnarray}}
\newcommand{\bse}{\begin{subequations}}
\newcommand{\ese}{\end{subequations}}
\newcommand{\Rv}{\mathbf{R}}
\newcommand{\upd}{\text{d}}

\newcommand{\bkfa}{Ba\textsubscript{1-x}K\textsubscript{x}Fe\textsubscript{2}As\textsubscript{2}}

\begin{document}

\author{Alexandru Golic}
\affiliation{Department of Physics, KTH Royal Institute of Technology, Stockholm SE-10691, Sweden}
\author{Egor Babaev}
\affiliation{Department of Physics, KTH Royal Institute of Technology, Stockholm SE-10691, Sweden}
\affiliation{Wallenberg Initiative Materials Science for Sustainability, Department of Physics, KTH Royal Institute of Technology, SE-106 91 Stockholm, Sweden}
\author{Johan Carlström}
\affiliation{Department of Physics, KTH Royal Institute of Technology, Stockholm SE-10691, Sweden}

\title{
 Demonstration of a fermion Quadrupling Condensate via Quantum Monte Carlo Simulation
}
\begin{abstract}
 
Fermionic condensation typically occurs via pairing. In recent decades, however, a fundamental question has emerged: whether alternative forms of order exist, such as condensates of fermion quadruplets. These states--including ``charge-4e" superconductors and ``charge-0" counterflow condensates--lie beyond the standard Bardeen-Cooper-Schrieffer framework, and require strong fluctuations and correlation effects that invalidate the BCS mean-field description. 
This makes the problem notoriously difficult to study numerically at a microscopic level, as it involves both strong interactions and the fermionic sign problem.
Here, we present a microscopic fermionic model featuring correlated hopping that significantly mitigates the sign problem, enabling rigorous Monte-Carlo-based analysis. Using large-scale simulations, we demonstrate the existence of a fermion-quadrupling condensate with a transition temperature comparable to the hopping energy scale. These results provide direct numerical evidence for quartic fermionic order in a microscopic system and suggest that these exotic states are also experimentally accessible in ultracold atomic gases.

\end{abstract}

\maketitle

 The electron pairing problem is subtle as electrons naturally repel each other in a vacuum. 
The attractive interaction therefore needs to be mediated by a medium, as famously described by the Bardeen-Cooper-Schriefer (BCS) theory \cite{Bardeen1957a,Bardeen1957b,Cooper1956}.  
That framework, however, does not allow for higher-order bound states; within BCS theory, electron quadruplets cannot win energetically or entropically over pairs, see \cite{wang2023origin} for further discussion.

Recent experiments on multicomponent systems \cite{Grinenko2021state,shipulin2023calorimetric,halcrow2024probing,Bartl2024} have significantly intensified theoretical investigations into the problem of electron condensation into higher-order composites. These studies suggest four-electron order, characterized by symmetry breaking driven by correlated electron quadruplets in the absence of conventional pair condensation.
 Such phases lie beyond the BCS paradigm, but have been predicted when fluctuations are included in multicomponent London and Ginzburg–Landau theories \cite{babaev2002phase,babaev2004superconductor,Bojesen2014phase,Maccari2022effects,Herland2010phase,agterberg2008dislocations,Radzihovsky2009liquid,berg2009charge}. More recently, a microscopic approach  was developed to address fermion quadrupling directly at the fermionic level within generalized mean-field treatments in the moderately weak-coupling regime \cite{samoilenka2025microscopic}.

In order to have a superconducting order one needs to have an attractive mechanism by which the fermions are bound into pairs.
However, with the proper binding mechanism there is nothing preventing condensates with an arbitrary (even) number of fermions; an obvious example of this is
superfluid He${}^4$, where each atom is a direct bound state of two electrons and four nucleons, themselves consisting of 12 quarks.
Recently related  types of directly bound multi-fermion objects have also attracted interest as an avenue to charge-$4e$ superconductors \cite{soldini2024charge,wan2026quantumcharge4esuperconductivitydeconfined}.
For condensates composed of four or more fermions, however, a distinct microscopic mechanism exists that can give rise to a four-fermion ordered state. If the system hosts a large population of preformed yet uncondensed pairs in different bands, four-fermion order may emerge purely from correlations in the motion of these pairs. This mechanism enables the condensation of higher-order flows, while single-component flows remain dissipative.
Correlated motion is ubiquitous in multicomponent systems and can arise from gauge-field-mediated coupling as well as from strong-correlation effects such as the Andreev–Bashkin drag.

This type of four-fermion order can arise in several distinct forms in multicomponent systems. One class corresponds to counterflow phases, where pairs in different components flow in opposite directions, resulting in a neutral superfluid mode without global phase coherence. A second class involves coflow of pairs,  in particular giving rise to charge-$4e$ superconductivity \cite{Radzihovsky2009liquid,berg2009charge,Herland2010phase,babaev2004superconductor}. An interesting realization occurs in systems with discrete symmetry breaking, such as time-reversal symmetry, where an intermediate phase retains composite order after superconductivity is lost, as reported in \bkfa \cite{Grinenko2021state}. These scenarios represent closely related manifestations of fermion quadrupling, distinguished by the symmetry of the underlying collective mode.

A key distinction between the mechanism considered here and direct particle binding or explicit quadrupling is most transparent in three-component counterflow. In this case, only pairs of different fermionic species can counterpropagate, yielding a four-fermion order parameter without a uniquely defined counterpropagating partner \cite{blomquist2021borromean,babaev2024hydrodynamics,golic2025borromean}. Accordingly, the appropriate description must incorporate quasiparticles, short-range incoherent pairing, and long-range four-fermion order.

Ultracold atoms in optical lattices provide a natural platform for realizing novel forms of composite order. In particular, significant progress has been made for counterflow phases in multicomponent bosonic systems, characterized by order parameters bilinear in the bosonic fields. Such bosonic phases were first predicted based on general arguments and Monte Carlo studies of classical field theories \cite{kuklov2003counterflow,altman2003phase,kuklov2004commensurate,kuklov2004superfluid}, and subsequently demonstrated in quantum lattice models \cite{soyler2009sign,Sellin18superfluiddrag,blomquist2021borromean}. Most recently, counterflow order has been realized experimentally in optical lattice systems \cite{Zheng2025}. This progress establishes ultracold atoms as a promising platform for exploring composite fermionic orders.

Theoretically, however, composite order phases are considerably more challenging to demonstrate for electronic system than their bosonic counterparts due to the fermionic sign problem.
Consequently, theoretical descriptions have largely relied on a two-step approach: first, constructing phenomenological Ginzburg-Landau or London models, and second, incorporating fluctuations that stabilize composite order beyond the mean-field level. These studies indicate that robust electron-quadrupling phases require strong or long-range correlations, making direct numerical treatment of fermionic Hamiltonians particularly challenging. This motivates the need for controlled numerical demonstrations of quadrupling order in microscopic fermionic systems, as well as the identification of minimal Hamiltonians that can realize such phases. At the same time ultracold fermionic atoms provide a promising quantum emulator platform  to realize such phases, offering tunable interactions, filling, and correlated hopping, and enabling multicomponent systems with independently conserved condensates and $U(1)\times U(1)$ symmetry \cite{bloch2008many,bloch2012quantum,lewenstein2012ultracold,altman2021quantum}.

In this work we construct a lattice Hamiltonian of a $U(1)\times U(1)$ fermionic system, and present a numerical demonstration of quadrupling states in a fermionic system, in a direct Monte-Carlo simulation. 
In contrast to previous analytical work, which aims at the moderately weak-coupling scenario \cite{samoilenka2025microscopic}, we focus on the strongly correlated regime. 
These results not only demonstrate the existence of higher-order condensates but also suggest their realizability in ultracold atomic gases in optical lattices.

{\it Model---} As discussed in the Introduction, classical field-theoretical studies indicate that fermion quadrupling can emerge in systems with strong current–current correlations, arising either from electronic interactions or from long-range intercomponent gauge-field coupling. Here we focus on the interaction-driven case and introduce a minimal two-component lattice model that provides a microscopic realization of such intercomponent current correlations:
\begin{align}
    \label{eq:hamiltonian}
    H = - t\sum_{\left\langle ij\right\rangle\alpha\sigma}\cd_{i\alpha\sigma}c_{j\alpha\sigma} + U\sum_{i\alpha}\hat{n}_{i\alpha\uparrow}\hat{n}_{i\alpha\downarrow} \\ \nonumber
    -d\sum_{\left\langle ij\right\rangle\alpha}p^{\dagger}_{i\alpha}p_{j\alpha} - q\sum_{\left\langle ij\right\rangle}p^{\dagger}_{i1}p^{\dagger}_{j2}p_{i2}p_{j1}
    - \mu\sum_{i\alpha\sigma}\hat{n}_{i\alpha\sigma}
    .
\end{align}
Here $i,j$ label lattice sites, $\alpha = 1,2$ denotes the component index, and $\sigma = \uparrow,\downarrow$ the spin. The operators $c_{i\alpha\sigma}$ are fermionic annihilation operators, and $p^{\dagger}_{i\alpha} = \cd_{i\alpha\uparrow}\cd_{i\alpha\downarrow}$ creates an on-site pair. The Hamiltonian is parametrized by the single-particle hopping amplitude $t$, the pair-hopping amplitude $d$, the correlated hopping amplitude $q$, the chemical potential $\mu$, and the intraband contact interaction $U$.

The physics we aim to capture is not strongly tied to the microscopic details of this Hamiltonian. Rather, the model serves as a minimal setting that realizes strong intercomponent correlations, representative of multicomponent systems far beyond the regime of validity of the BCS mean-field approximation.

The Hamiltonian (\ref{eq:hamiltonian}) may be viewed as a generalization of the Penson–Kolb–Hubbard model \cite{pensonkolb1986original, robaszkiewicz1999pair}, which was proposed as an effective description of superconductivity in strongly correlated systems.
In contrast to that model, our description involves two fermionic components and includes intercomponent correlated pair hopping. On mesoscopic scales, this term induces intercomponent drag—a phenomenon also present in two-component Fermi liquids \cite{Sjoberg:76}, where it is particularly strong in dense nuclear matter and is believed to play an important role in pulsar dynamics \cite{chamel2013crustal, borumand1996neutron, chamel2008twosuperfluid}.

While the model (1) exhibits drag that favors counterflow, applying particle-hole transformation to one of its bands results in a dual description that promotes coflow of pairs. In the absence of a gauge field, we can thus treat both these scenarios on an equal footing.

We focus on the strong-coupling regime, and hence
 choose the following model parameters:
\begin{align}
    \label{modelParameters}
    \mu=-3.1,\; U=-6,\; t=1,\; d=1,\; q=3.
\end{align}

 In contrast to bosonic pairing, fermionic quadrupling has long eluded numerical treatment due to the fermionic sign problem, which generically leads to an exponentially decaying signal-to-noise ratio with increasing system size \cite{troyer2005signproblem}. This severely limits accessible system sizes, rendering them insufficient to resolve continuous phase transitions in two dimensions—the central goal of this work. 
However, in this model, it is possible to overcome this limitation because here the fermionic exchange is suppressed: we work in a regime where fermions form moderately bound pairs, and employ a hexagonal lattice that geometrically suppresses exchange of hard-core particles.

Even in this regime, the sign problem is sufficiently severe to preclude brute-force fermionic simulations of all aspects of the phase transitions. However, a large sign problem does not necessarily imply that Fermi statistics strongly affect all observables. For certain observables and parameter ranges, expectation values can depend only weakly on the fermionic sign if exchange processes-the origin of the sign-do not contribute significantly. For such observables, we can, as explained below, with a controlled accuracy, effectively ``turn off" the fermionic sign in the simulations (equivalent to replacing the four fermionic species with hard-core bosons), thereby accessing much larger system sizes.
To validate the approach, we simulate both the sign-free model, to access the system sizes required to probe the transitions, and the full fermionic model, to validate this procedure. Further discussion of the sign problem is provided in the Appendix.

{\it Superfluid stiffness}---We characterize the phases of model (\ref{eq:hamiltonian}) via the phase-stiffness matrix. In the ground state, the U(1)$\times$U(1) symmetry is expected to be fully broken, yielding two independently conserved condensates that flow without dissipation. In this regime, the hydrodynamic/hydrostatic description is, by symmetry given by a classical field theory involving two phases, $\phi_i$, corresponding to the two types of pairs, with the free energy given by
\begin{align}
    \label{eq:effective-long-range-single-drag}
    F[\phi] = \frac{1}{2T}\int\upd^2r\hspace{3pt} \rho_1(T)\sum_i|\nabla\phi_i|^2 + 2\rho_d(T)\nabla\phi_1\cdot\nabla\phi_2.
\end{align}
Here $\rho_1(T)$ is the (bare) superfluid stiffness of the individual phases, and $\rho_d(T)$ the (bare) Andreev-Bashkin drag \cite{Andreev1975}, which--being permitted by symmetry--is necessarily present when the components interact.
To elucidate the quadrupling transition, we recast the free energy (\ref{eq:effective-long-range-single-drag}) as
\begin{align}
    \label{eq:effective-long-range-co-counter}
    F[\phi] = \frac{1}{2T}\int\upd^2r\hspace{3pt} \frac{\rho_{(1, 1)}}{4}| \sum_i \nabla\phi_i|^2 \!+\! \frac{\rho_{(1,-1)}}{4}|\nabla [\phi_1\!-\!\phi_2]|^2,
\end{align}
where $\rho_{(1, \pm1)} = 2(\rho_1 \pm \rho_d)$ are the (bare) co- and counterflow stiffnesses, respectively.
In the context of the classical field theory (\ref{eq:effective-long-range-co-counter}), the aforementioned particle-hole transformation of the second band corresponds to a phase inversion $\phi_2\to -\phi_2$. This is equivalent to interchanging 
$\rho_{(1, 1)} \leftrightarrow \rho_{(1, -1)}$.

For the free energy to be bounded from below, both composite stiffnesses must be positive, implying $|\rho_d| \leq |\rho_1|$.
It is therefore useful to consider the relative drag
\begin{align}
    \rho_r\equiv    \frac{\rho_d}{\rho_s} = \frac{\rho_{(1, 1)} - \rho_{(1, -1)}}{\rho_{(1, 1)} + \rho_{(1, -1)}},\label{relDrag}
\end{align}
whose magnitude remains bounded $|\rho_r| \leq 1$.

At finite temperatures the long-range interactions get screened by vortex-antivortex pairs leading to an effective scale dependence of the superfluid stiffnesses
and Andreev-Bashkin drag, such that $\rho_t = \rho_t(L)$ for $t = s, d, (1,\pm1)$ (although the relationship $\rho_{(1, \pm1)}(L) = 2(\rho_1(L) \pm \rho_d(L))$ still holds for all scales).
In this formulation order (disorder) in a sector corresponds to a finite (vanishing) stiffness $\rho(\infty)$ for infinite system size.
The flow of the stiffnesses as a function of system size is described by a generalization of Kosterlitz-Thouless RG, which we derive in detail in the Appendix.

At zero drag, the condensates decouple, and the restoration of $U(1)\times U(1)$ symmetry is driven by the proliferation of single-component vortices that exhibit winding only in one of the phases.
With increasing drag, one of the composite stiffnesses
$\rho_{(1, \pm1)}(L)$ is reduced while the other is increased. 
When the drag becomes sufficiently large such that $\rho_{(1, \pm1)} < \rho_1$ (i.e. for relative drag $|\rho_r| > 0.5$) the critical behavior changes qualitatively, and the transition from the superfluid phase is instead driven by the proliferation of composite vortices with phase winding in $\phi_1 \pm \phi_2$. For a detailed derivation of the critical behavior see the Appendix.

When the relative drag is close to $\rho_r=0.5$, this transition may still destroy the order in the complementary sector $\phi_1 \mp \phi_2$, thereby restoring the full $U(1)\times U(1) $ symmetry in a single phase transition. 
For sufficiently large microscopic drag, however, the complementary sector can remain ordered, i.e. one of $\rho_{(1, \pm1)}(\infty)$ vanishes in the infinite system while the other remains finite (with $|\rho_r(\infty)| = 1$). 
In this case, the remaining order vanishes only at a higher temperature; the resulting intermediate phase represents a quadrupling state, that is described by a four-fermion order parameter. The nature of this phase depends on the sign of the drag term: For $\rho_d<0$, the system can transition to a counterflow superfluid (i.e., order only in the phase difference composed field $\phi_1-\phi_2$). For $\rho_d>0$, the system can transition to a state with order only in the phase sum composed field $\phi_1+\phi_2$),
In a gauged system, that state has doubled coupling to the vector potential and hence represents a charge-$4e$ superconductor.

{\it Numerical Computations}---We compute the superfluid response of the microscopic model (\ref{eq:hamiltonian}) by sampling imaginary-time path integrals at thermal equilibrium using the Worm algorithm. The scale dependent superfluid stiffnesses and Andreev-Bashkin drag are extracted from winding-number fluctuations via a generalization of the Pollock–Ceperley formula \cite{polock1987winding} to the multicomponent action (\ref{eq:effective-long-range-co-counter}). In two dimensions, this estimator is sensitive to the system aspect ratio \cite{prokofev2000twodefinitions}, and we therefore employ approximately square geometries.
For square lattices, the linear system size is typically defined as the number of sites along one direction. For the honeycomb lattice, we instead define $L=\sqrt{N}$, where $N$ is the total number of sites.

In two dimensions, superfluid transitions are of the Berezinskii–Kosterlitz–Thouless (BKT) type. 
The critical temperatures are determined from finite-size scaling of the superfluid stiffness \cite{polock1987winding}.
For a finite system of linear size $L$, the scale-dependent critical temperature $T_c(L)$ is defined by the condition that the relevant scale-dependent superfluid stiffness $\rho(L,T)$ reaches the universal value $\rho_c$. For BKT transitions, $\rho_c=2T/\pi$ for the superfluid–quadrupling transition and $\rho_c=8T/\pi$ for the quadrupling–normal transition (see Appendix).
For large enough system sizes, the value of $T_c(L)$ follows the universal scaling law 
\begin{align}
    \label{eq:bkt_scaling}
    T_c(L) = T_{\text{BKT}} + \frac{A}{(\ln L/L_0)^2},
\end{align}
where $T_{\text{BKT}}$ is the critical temperature of the infinite system, and $A, L_0$ are non-universal constants.

For each system size and temperature we run multiple independent simulations over several autocorrelation times and from this calculate the mean and variance of the various superfluid stiffnesses $\rho(L, T)$.
To estimate $T_c(L)$ we fit a temperature sweep of these curves using the Pool Adjacent Violators Algorithm(PAVA) to find where the curves of $\rho(L, T)$ intersect $\rho_c$,
with the error being estimated by resampling the various $\rho(L, T)$ using the bootstrap method to get several samples of $T_c(L)$.
Finally the fit of $T_c(L)$ to the universal scaling form is made using the SCIPY function curve\_fit which also estimates the error of the parameters.

\begin{figure}[!htb]
    \includegraphics[width=\linewidth]{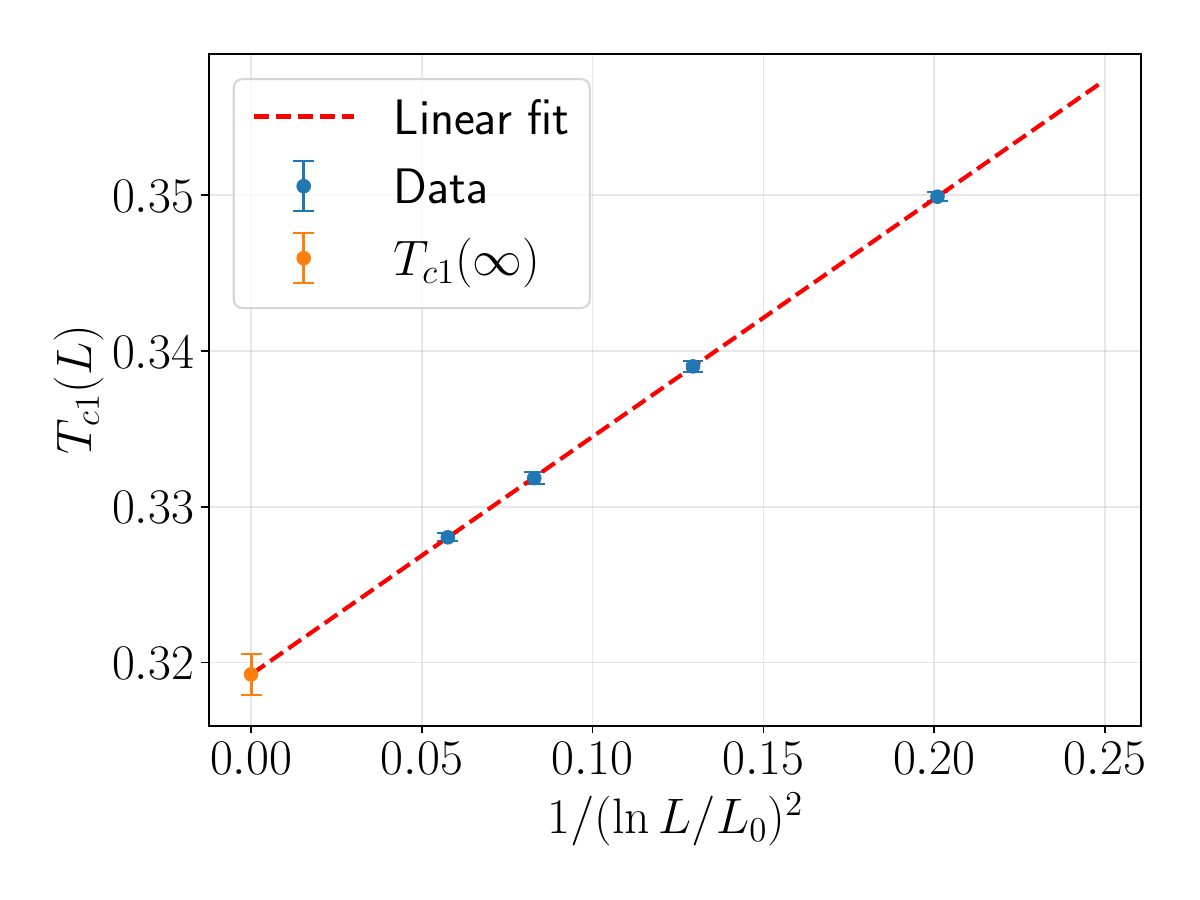}
    \caption{Critical temperature of the superfluid-quadrupling transition for the sign-free simulations as a function of $1/(\ln L/L_0)^2$ where $L$ is the linear system size and $L_0$ is a fit parameter from the finite-size scaling. 
    Points show simulation data with error bars and the solid lines represent the linear fit to the universal scaling form.
    This fit corresponds to the parameters $T_{c1} \approx 0.3192 \pm 0.0013$ (equivalent to $\beta_{BKT} \approx 3.133$), $A \approx 0.1525 \pm 0.0376$ and $L_0 \approx 4.25 \pm 0.98$.}
    \label{fig:sf_scaling}
\end{figure}

{\it Results}---To test the existence of a   quadrupling phase, we monitor the phase stiffness as a function of temperature and system size for both fermions and bosons, using the model parameters (\ref{modelParameters}). 
For the benchmark case of bosonic statistics, we identify two phase transitions: 
At a lower $T_{c1}\approx 0.32$, co-flow superfluidity is destroyed, leaving behind a counter-flow superfluid, see Fig. \ref{fig:sf_scaling}. At a higher critical temperature of $T_{c2}\approx 0.81$, this second superfluid mode is destroyed, restoring the full $U(1)\times U(1)$ symmetry. The intermediate phase is described by a four particle order parameter of the form $O\sim \langle c^\dagger_{\alpha}c^\dagger_{\alpha} c_{\beta}c_{\beta} \rangle$. 

The fermionic case is computationally more challenging due to the sign problem, and the phase diagram must be inferred from multiple observables and smaller system sizes. Figure 5 shows the relative difference in stiffness between bosons and fermions for system sizes up to 
$L=34$. Below $T_{c1}$, the fermionic-model stiffness is $3-4\%$ lower than in the bosonic case, suggesting correspondingly a lower critical temperature in the thermodynamic limit. In contrast, near $T_{c2}$  the discrepancy is much smaller, 
$0.1-0.4\%$ indicating that Fermi statistics plays a reduced role at higher temperatures, where space-time paths are shorter and particle exchanges are suppressed. This implies that the intermediate quadrupling regime is broader for fermions than for bosons. 

Notably, in the limit $U\to-\infty$, the fermionic and bosonic models coincide. In this case, the upper critical temperature is pushed to $T_{c2}\approx 0.97$ due to the suppression of single-particle fluctuations (see Appendix III). Thus, while particle statistics has only a minor effect on $T_{c2}$, single-particle fluctuations play a significant role.

Finally, we investigate the temperature dependence of the relative drag—defined in Eq.~(\ref{relDrag})—at a fixed system size of $L\approx68$, 
while varying the correlated hopping amplitude $q$ of the model (\ref{eq:hamiltonian}), in an interval $1\le q\le 3$ (see Fig.~\ref{fig:drag}).
For parameters where the transition from the ordered states is driven by the proliferation of single vortices ($|\rho_r| < 0.5$) 
the relative drag $\rho_r$ flows towards zero with increasing temperature, while for parameters where the transition is driven by the proliferation of composite vortices
($|\rho_r| > 0.5$) it flows to unity. 

\begin{figure}[!htb]
    \includegraphics[width=\linewidth]{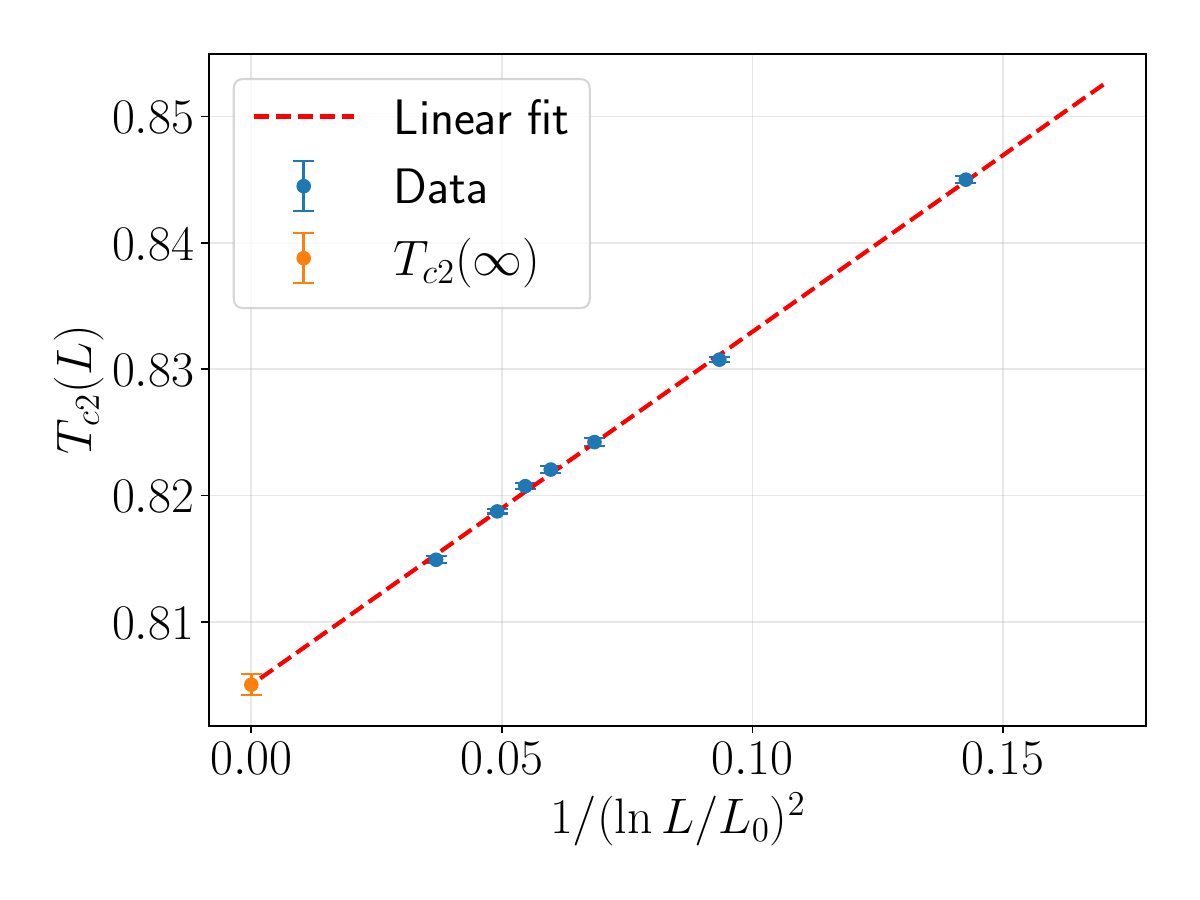}
    \caption{Critical temperature of the quadrupling-normal transition for the sign-free simulations as a function of $1/(\ln L/L_0)^2$, where $L$ is the linear system size and $L_0$ is a fit parameter from the finite-size scaling. 
    Points show simulation data with error bars and the solid lines represent the linear fit to the universal scaling form.
    This fit corresponds to the parameters $T_{c2} \approx 0.805 \pm 0.001$ (equivalent to $\beta_{BKT} \approx 1.242$), $A \approx 0.275 \pm 0.032$ and $L_0 \approx 1.52 \pm 0.20$.}
    \label{fig:quad_scaling}
\end{figure}

\begin{figure}
    \centering
    \includegraphics[width=\linewidth]{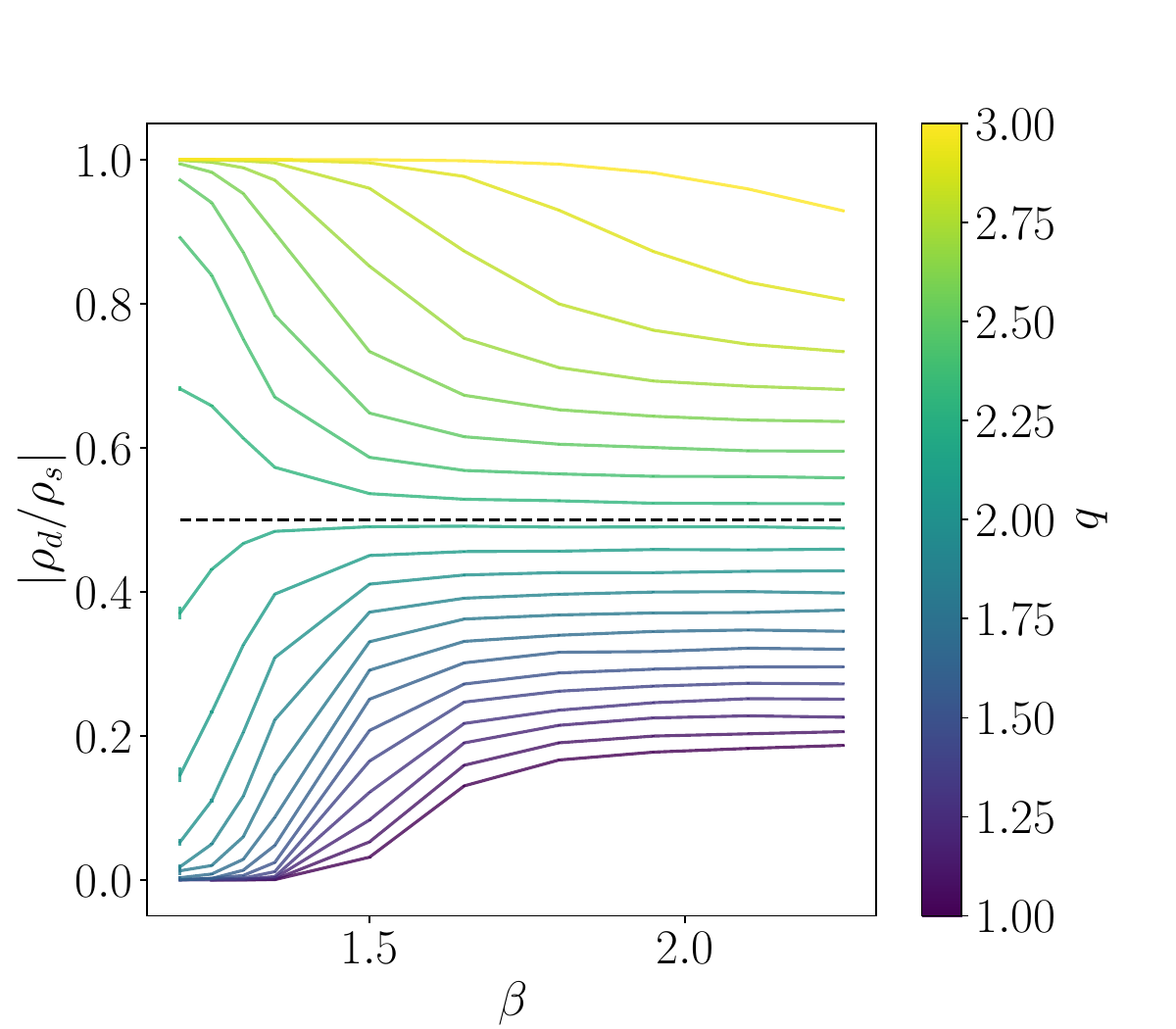}
    \caption{Relative intercomponent drag $|\rho_d/\rho_s|$ as a function of inverse temperature $\beta$ and microscopic counter-hopping strength $q$ at the system size $L \approx 68$ for the sign-free model.
    The dashed line denotes $\rho_d/\rho_s = 0.5$.
    The high temperature behavior of the drag for each $q$ signals the critical nature of the transition away from superfluidity. For $q \lesssim 2.25$ the transition is driven by the disordering of single phases $\phi_1, \phi_2$, while for $q \gtrsim 2.25$ it is driven by disordering of one of the two linear combinations $\phi_1 \pm \phi_2$.
    The line $|\rho_d/\rho_s| = 0.5$ is the marginal point where the single component superfluid stiffness is equal to one of the composite superfluid stiffnesses.
    }
    \label{fig:drag}
\end{figure}

\begin{figure}
    \centering
    \includegraphics[width=\linewidth]{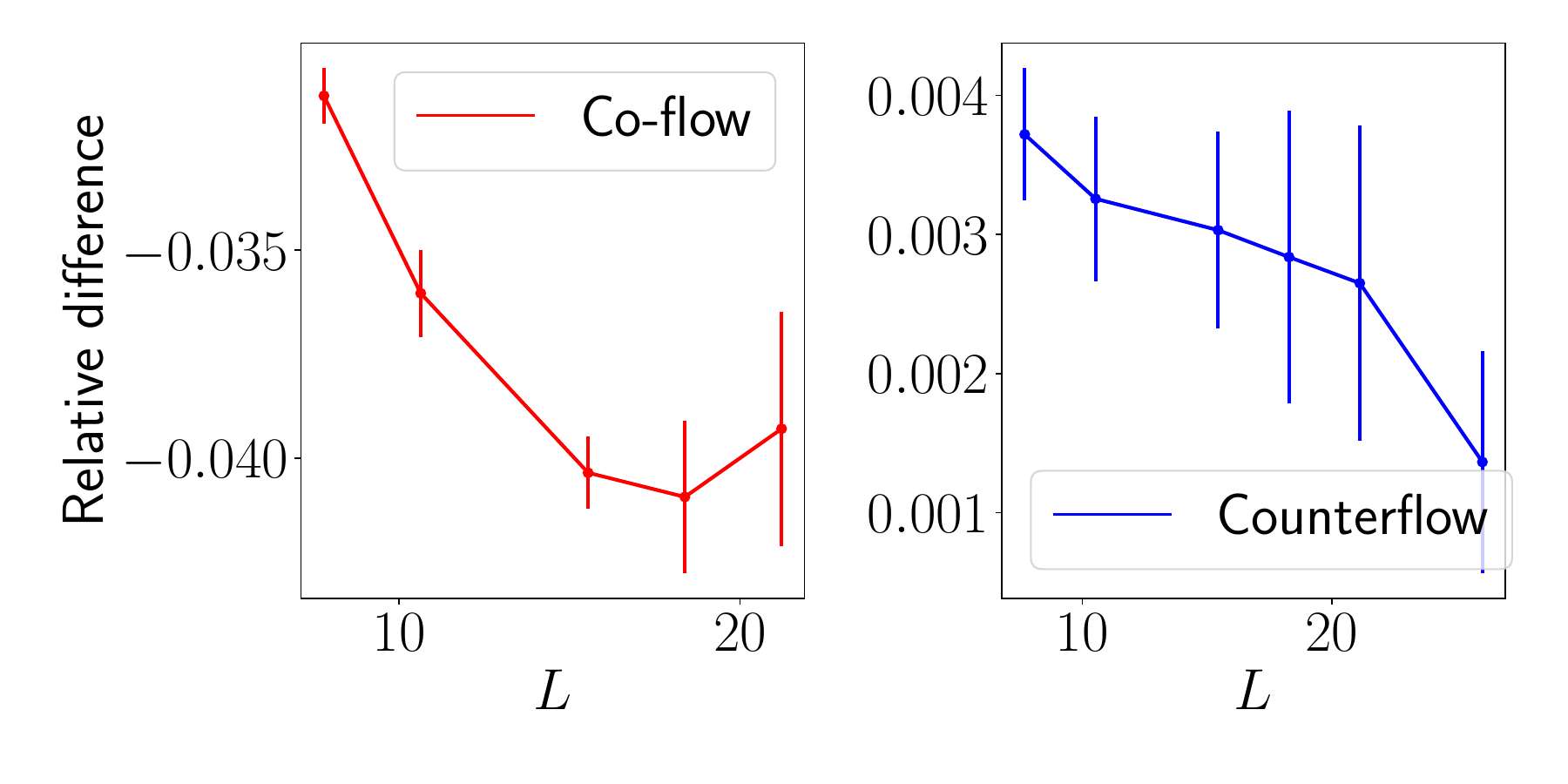}
    \caption{Relative difference in phase stiffness $\rho_{\text{Fermi}}/\rho_{\text{Bose}} - 1$ between bosons and fermions in a system with negative drag. 
    The left picture shows the relative difference in the co-flow stiffness at $T = 0.25$, significantly below the superfluid-supercounterfluid transition for the bosonic system.
    The right picture shows the relative difference in the counter-flow stiffness at $T \approx 0.77$, slightly below the  supercounterfluid-normal transition for the bosonic system.}
    
    \label{fig:sign-comp}
\end{figure}

{

{\it Discussion}---We have presented a microscopic demonstration of a fermion-quadrupling condensate in a fermionic lattice model. This establishes that higher-order fermionic order can emerge directly from interacting fermions in a controlled setting, without relying on phenomenological assumptions.

In contrast to recent analytical work focused on the moderately weak-coupling regime \cite{samoilenka2025microscopic}, our approach addresses the strongly correlated limit. Unlike the BCS paradigm—where condensation is governed by coherent pair formation—we find that quadrupling order arises from correlated hopping. In this regime, quasiparticles and short-range pairs persist, but the pairs themselves do not exhibit long-range coherence. Instead, algebraic order appears only in four-fermion correlators, signaling a genuine fermion-quadrupling phase.

Although in the present model the effective drag originates from correlated pair hopping, the underlying mechanism is more general and can arise from a broad class of interaction-induced correlation effects in multicomponent systems. Our results demonstrate both counterflow quadrupling order,
$O\sim \langle c^\dagger_{\alpha}c^\dagger_{\alpha} c_{\beta}c_{\beta} \rangle$ and charge-$4e$ superconductivity,
$O\sim \langle c_{\alpha}c_{\alpha} c_{\beta}c_{\beta} \rangle$, which are related by a duality transformation. 

These findings provide a concrete route toward realizing fermion-quadrupling condensates in experimentally accessible platforms such as ultracold fermionic mixtures in optical lattices. More broadly, they show that strong correlations can fundamentally alter the structure of fermionic condensation, enabling composite orders beyond the conventional pairing paradigm.
}

\section*{Acknowledgments}
We thank Albert Samoilenka for discussions.
We acknowledge support from the project grant KAW 2024.0131 from Knut och Alice Wallenbergs Stiftelse,
AG and EB were supported by the Swedish Research Council Grants  2022-04763, 
 EB was  partially supported by the Wallenberg Initiative Materials Science
for Sustainability (WISE) funded by the Knut and Alice Wallenberg
Foundation.
JC was supported by  Olle Engkvists stiftelse
through Grant No. 240-0803.
The simulations 
were enabled by resources provided by the National Academic Infrastructure for Supercomputing in Sweden (NAISS), 
partially funded by the Swedish Research Council through grant agreement no. 2022-06725.

\bibliography{LITTLEBIBFILE}

\appendix

\subsection{Appendix I: RG equations} \label{app:rg}
While the microscopic model that we simulate is fermionic, by symmetry its macroscopic response is governed by two coupled phase fields and the phase transitions are governed by proliferation of topological defects in these fields. 
Here we will present the derivation of RG equations in brief, for a more detailed review of the standard Kosterlitz-Thouless RG consult e.g \cite{kosterlitz1973rg, Svistunov2015}.

At long distances our model of two superfluids can be described by an effective model of two condensates with phases $\phi_1, \phi_2$ that interact through Andreev-Bashkin drag. 
Since the model is component-symmetric, the resulting free energy can be written as
\begin{align}
    F[{\phi_i}] = \frac{1}{2}\int\text{d}^2r\hspace{5pt} \Lambda_s(|\nabla\phi_1|^2+|\nabla\phi_2|^2) + 2\Lambda_d\nabla\phi_1\cdot\nabla\phi_2, 
\end{align}
where $\Lambda_s = \rho_s/T$ is the unitless single component superfluid stiffness and $\Lambda_d = \rho_d/T$ is the unitless drag.
Single vortices (i.e (1, 0) or (0, 1)) interact with each other with the strength $\Lambda_s$,
while composite vortices with charges $(1, \pm1)$ interact with each other with the strength $\Lambda_{(1, \pm1)} \equiv 2\Lambda_s \pm 2\Lambda_d$.
In our system the drag is significant with a magnitude comparable to $\Lambda_s$.
For definiteness let us assume that $\Lambda_d$ is negative and favors counterflows (the analysis for a co-flow state will follow the exact same procedure, so this assumption is w.l.g). 
Therefore the three lowest energy pairs are (1,0), (0,1) and (1,1) vortex pairs. 

In the presence of phase twists $\varphi_{1,2}$ along the $x$-axis in both components, these vortex pairs have the energies
\begin{align}
    &E_{(1, 0)}(\Rv) = 2\pi\Lambda_s\ln{|\Rv|} - \frac{\pi}{L_x}(2\Lambda_s\varphi_1 - 2\Lambda_d\varphi_2)\xh\cdot\left(\zh\times\Rv\right), \\
    &E_{(0, 1)}(\Rv) = 2\pi\Lambda_s\ln{|\Rv|} - \frac{\pi}{L_x}(2\Lambda_s\varphi_2 - 2\Lambda_d\varphi_1)\xh\cdot\left(\zh\times\Rv\right), \\
    &E_{(1, 1)}(\Rv) = 2\pi\Lambda_{(1, 1)}\ln{|\Rv|} - \frac{\pi}{L_x}\Lambda_{(1, 1)}(\varphi_1 + \varphi_2)\xh\cdot\left(\zh\times\Rv\right).
\end{align}

Assuming that the vortex pairs are well separated from each other (i.e that the vortex pairs effectively form a dilute gas) the free energy for a system of size $L$ can be written as
\begin{equation}
    F(L) = (L/l_0)^2F(l_0) - V\int_{l_0 \leq \Rv \leq L}\left(e^{E_{(1, 0)}} + e^{E_{(0, 1)}} + e^{E_{(1, 1)}}\right)\upd^2 R,
\end{equation}
where $l_0$ is a small size cut-off and $V$ is the system size.

Letting $8\pi g_s = e^{(1, 0)} = e^{(0, 1)}$ and $8\pi g_{(1, 1)} = e^{E_{(1, 1)}}$ we can rewrite this as
\begin{equation}
    F(L) = (L/l_0)^2F(l_0) - \frac{V}{8\pi}\int_{l_0 \leq \Rv \leq L}2g_s + g_D\upd^2 R.
\end{equation}
Taking the derivative w.r.t the phase twist in the first component at zero supercurrents we get
\begin{align}
    &\Lambda_s'(l) = -\frac{\pi^2}{4}(4\Lambda_s^2 + 4\Lambda_d^2)g_s - \frac{\pi^2}{4}\Lambda_{(1, 1)}'(l)^2g_D, \\
    &\Lambda_{(1, 1)}'(l) = -2\frac{\pi^2}{4}\Lambda_{(1, 1)}^2(l)^2g_s -4\frac{\pi^2}{4}\Lambda_{(1, 1)}^2(l)^2g_D .
\end{align}

Noting how $g_{s,(1, 1)}$ scale with $L$ we get
\begin{align}
    &\Lambda_s'(l) = -\frac{\pi^2}{4}(8\Lambda_s^2 - 4\Lambda_s\Lambda_{(1, 1)} + \Lambda_{(1, 1)}^2)g_s - \frac{\pi^2}{4}\Lambda_{(1, 1)}^2g_{(1, 1)},\\
    &\Lambda_{(1, 1)}'(l) = -2\frac{\pi^2}{4}\Lambda_{(1, 1)}^2g_s -4\frac{\pi^2}{4}\Lambda_{(1, 1)}^2g_{(1, 1)}, \\
    &g_s' = -2(\pi\Lambda_s - 2)g_s, \\
    &g_{(1, 1)}' = -2(\pi\Lambda_{(1, 1)} - 2)g_{(1, 1)}.
\end{align}

Note now the following: Because (1, 0), (0, 1) and (1, 1) vortex pairs have the lowest energy, as increase temperature the two-superfluid phase will be destroyed either through the proliferation of (1, 0) and (0, 1) vortex pairs or by proliferation of (1, 1) vortex pairs.
Since all $\Lambda$'s are coupled to the density of single vortex pairs, if the single vortices proliferate first then all other vortices automatically also proliferate, and the whole system undergoes a single transition directly to the normal state.

If instead the (1, 1) vortices proliferate first, it is conceptually useful to rewrite the RG equations using in terms of the counter-flow stiffness (or equivalently the energy cost of a (1, -1) vortex) through $4\Lambda_s = \Lambda_{(1, 1)} + \Lambda_{(1, -1)}$ giving
\begin{align}
    \label{eq:rg_co_counter}
    &\Lambda_{(1, 1)}'(l) = -\pi^2\Lambda_{(1, 1)}^2\left(g_{(1, 1)} + g_s/2\right), \\
    &\Lambda_{(1, -1)}'(l) = -\frac{\pi^2}{2}\Lambda_{(1, -1)}^2g_s, \\
    &g_s' = -2(\frac{\pi\Lambda_{(1, 1)}}{4} + \frac{\pi\Lambda_{(1, -1)}}{4} - 2)g_s, \\
    &g_{(1, 1)}' = -2(\pi\Lambda_{(1, 1)} - 2)g_{(1, 1)}.
\end{align}

After the (1, 1) vortices proliferate the co-flow stiffness $\Lambda_{(1, 1)}$ will be renormalized to zero, and the RG equations simplify to
\begin{align}
    \label{eq:rg_counter}
    &\Lambda_{(1, -1)}'(l) = -\frac{\pi^2}{2}\Lambda_{(1, -1)}^2g_s, \\
    \label{eq:rg_counter_density}
    &g_s' = -2(\frac{\pi\Lambda_{(1, -1)}}{4} - 2)g_s, \\
\end{align}
which are almost identical to the standard Kosterlitz-Thouless RG equations except for the factor $1/4$ in equation \ref{eq:rg_counter_density}.
This means that the critical value of $\Lambda_{(1,-1)}$ is shifted to
\begin{align}
    \frac{\pi\Lambda_c}{4} - 2 = 0 \implies \Lambda_c = 8/\pi,
\end{align}
but otherwise it acts exactly like a standard single-component model.

From these equations we can see that the RG flow of the vortex densities $g$ is fully determined by the relevant superfluid stiffness $\Lambda$.
If $\Lambda_s > \Lambda_{(1, 1)}$, then the number of single vortices will be exponentially suppressed with $l$ compared to the $(1, 1)$-vortices and vice versa.
This means that the critical nature of the system will be highly sensitive to which type of vortex is energetically cheapest.

For $\Lambda_s < \Lambda_{(1, 1)}$ the system is driven by the proliferation of single vortices.
Note however that the RG flow of $\Lambda_{(1, 1)}$ is coupled to the single vortex concentration $g_s$ in such a way that a proliferation
of single vortices ($g_s \to \infty$) also leads to an infinite suppression of $\Lambda_{(1, 1)}$ and thus a proliferation of these as well 
(which is to be expected, since a proliferation of all single vortices should lead directly to the normal state).

For $\Lambda_{(1, 1)} > \Lambda_s$ the system is instead driven by the proliferation of (1, 1) vortices.
In this case we also see that the RG flow of single vortices is coupled to double vortices but
in such a way that a proliferation of $(1, 1)$ vortices does not necessarily lead to a proliferation
of the single vortices.
This means that when (1, 1) vortices proliferate first there are two possible outcomes:
Either the resulting jump in the (1, -1) stiffness is large enough to proliferate them as well
giving us a single transition from superfluid to normal, or the jump is not large enough to proliferate them in which case the resulting state is a counter-flow state where single and (1, -1) vortices remain unproliferated.

At the marginal point $\Lambda_s = \Lambda_{(1, 1)}$ between these two cases corresponds precisely to the point where the relative drag 
\begin{align}
    |\rho_d/\rho_s| = \frac{\left|\Lambda_{(1, 1)} - \Lambda_{(1, -1)}\right|}{\Lambda_{(1, 1)} + \Lambda_{(1, -1)}} = 0.5.
\end{align}

If the counter-fluid state exists and the transitions between superfluid-counterfluid and counterfluid-normal are well separated it is easy to distinguish them.
In this case, the superfluid-counterfluid transition can be determined using the standard BKT scaling for the co-flow stiffness while ignoring the counter-flow stiffness,
while the counterfluid-normal transition can be determined using the BKT scaling form for the counter-flow stiffness (with the modified critical value $\Lambda_c = 8/\pi$)
under the condition that $\Lambda_{(1,1)} \approx 0$, to ensure that the two scales do not mix.

\section{Appendix II: Sign-problem}\label{app:sign}
When studying fermions with Monte Carlo we generically encounter the infamous "sign-problem".
This problem generically appears in Monte Carlo for all systems which have negative weights for the partition function.
For a system that does not suffer from the sign problem, the partition function is simply given by a sum of positive weights
$\mathcal{Z} = \sum_n w_n$, $w_n > 0$ $\forall n$, with the expectation value for an observable $\mathcal{O}$ being given by
\begin{align}
    \braket{\mathcal{O}} = \frac{1}{Z}\sum_n w_n\mathcal{O}_n,
\end{align}
where $\mathcal{O}_n$ is the value of the observable for the state $n$.
This can straightforwardly be simulated using normal Monte Carlo methods by interpreting the weights $w_n/\mathcal{Z}$ as probability amplitudes.

If the weights in the partition function are allowed to be negative, then the standard way of handling this is by treating the magnitude of the weights $|w_n|$
as probability amplitudes together with the "sign-free" partition function $\mathcal{Z}_+ \equiv \sum_n |w_n|$ such that the probability of being in state $n$ is given by $|w_n|/\mathcal{Z}_+$.
Writing $w_n = \sigma_n|w_n|$, where $\sigma_n = \pm1$, this allows us to write the true partition function using the expectation value of the sign and the sign-free partition function
\begin{align}
    \mathcal{Z} = \sum_n w_n = \sum_n |w_n| \sigma_n = \braket{\sigma}_+\mathcal{Z}_+,
\end{align}
where the subscript $+$ for the bracket indicates the average for sign-free weights. 
Similarly, the expectation value for any observable $\mathcal{O}$ can be written in terms of the sign-free partition function as
\begin{align}
    \braket{\mathcal{O}} = \frac{\sum_n w_n \mathcal{O}_n}{\mathcal{Z}} = \frac{\sum_n |w_n| \sigma_n \mathcal{O}_n}{\braket{\sigma}\mathcal{Z}_+} = \frac{\braket{\sigma\mathcal{O}}_+}{\braket{\sigma}_+}
\end{align}

While all Monte Carlo simulations suffer from numerical noise that increases with system size the scaling 
for a model with negative weights is much worse than for positive-definite weights
since effectively the calculation of $\braket{\mathcal{\sigma\mathcal{O}}_+}/\braket{\sigma}_+$ requires very precise
cancellation between the positive and negative sign states.
This leads to an exponential scaling of the noise with system size, as opposed to typically polynomial scaling for positive-definite weights \cite{troyer2005signproblem}.
This effectively sets an upper limit on the system sizes and temperatures which can directly be studied using Monte Carlo in systems that suffer from the sign problem.

The severity of the sign problem can vary significantly depending on the particular system we are simulating.
In the case of bosons, sign problems arise only in the case of certain repulsive interactions
but in the case of fermions the sign of the state equals $(-1)^{N_{\text{exchange}}}$, where $N_{\text{exchange}}$ is the number
of pair-wise exchanges of fermions within the imaginary-time periodic pat, and is thus generically present in fermionic systems.

Since the sign-problem arises due to exchanges of single fermions the sign-problem is however greatly diminished in systems where they are suppressed.
In this work we intentionally designed the model to minimize the sign-problem, both through the choice of a hexagonal lattice (which
requires at least 6 jumps in order to exchange the position of two fermions) and by working in the strong coupling limit where the physics is dominated by pairs of fermions, which have bosonic statistics.
Despite this the maximum tractable system sizes was around $N = 448, L \approx 21$ at $\beta = 4$ while for $\beta \leq 3$ $N = 680, L \approx 26$, both of which are far to small to resolve BKT transitions.

A subtle aspect however is that a large "sign-induced" noise does not always imply a large \textit{physical} effect.
For particular observables in certain parameter ranges it is possible for the effect of the sign to be negligible, even as the numerical simulations suffers from sign-induced noise. 
Within the framework of Path Integral Monte Carlo there is a very simple way of quantifying the physical effect of the sign separately from the numerical noise that it generates.
Since sign-full expectation values $\braket{\mathcal{O}} = \braket{\sigma\mathcal{O}}_+/\braket{\sigma}_+$ using the sign-free partition function, it is trivial to compute the sign free expectation value $\braket{\mathcal{O}}_+ = \mathcal{Z}_+^{-1}\sum_n |w_n|\mathcal{O}_n$ which is equivalent to replacing the fermions in our model with hard-core bosons.
For system sizes and temperatures where the sign-full problem is tractable we can then easily extract the relative difference $\delta_{\text{sign}}= \braket{\mathcal{O}}/\braket{\mathcal{O}}_+ - 1$ between having the sign turned on and off for the observable $\mathcal{O}$.
If the relative difference $\delta_{\text{sign}}$ is small it indicates that fermionic exchange processes contribute very little to the expectation value $\braket{\mathcal{O}}$ and conversely
a large $\delta_{\text{sign}}$ indicates that fermionic exchange processes are important for $\braket{\mathcal{O}}$.

The crucial aspect that we point out is: while we can only straightforwardly calculate the true fermionic expectation value $\braket{\mathcal{O}}$ for systems small enough to have a manageable sign problem, the quantity $\braket{\mathcal{O}}_+$ (which by construction does not suffer from the sign problem) can be accurately calculated for much larger systems (in our case, roughly two orders of magnitude more lattice sites for the superfluid stiffnesses).
Thus the sign-free simulations can essentially be seen as an approximation for the true sign-full problem which greatly reduces the computational complexity of the problem.
Furthermore this approximation is fully controlled for small system sizes since as mentioned previously it is trivial to extract the relative difference $\delta_{\text{sign}}$ between the sign-full and sign-free simulations.
If $\delta_{\text{sign}}$ is negligible for a given observable $\mathcal{O}$ at small system sizes then this strongly suggests that the expectation value is insensitive to the fermionic sign and thus justifies using this approximation for even larger system sizes.

It is worth emphasizing that this is {\it not} a general approximation valid for the whole system, but something that is particular for each observable.
In this paper we are interested in quantifying the fermionic quadrupling phase, which depends on the values of the two composite superfluid stiffnesses around the temperature range of the quadrupling phase.
Since fermions by themselves cannot condense these quantities necessarily depend on the motion of fermionic pairs, which makes them natural candidates for the aforementioned procedure of turning the sign off.
On the right side of figure \ref{fig:sign-comp} we can see the relative difference in the values of the larger composite stiffnesses (i.e. the counter-flow stiffness for negative drag and co-flow stiffness for positive drag) at $\beta = 1.3$ (slightly below the estimated sign-free quadrupling-normal transition) in turning the sign on or off.
Here we see that the fermionic sign has a very small effect ($<$1\% for the simulated range, and seemingly decreasing with system size) on the value of the stiffness,
meaning that the sign-free simulations are an accurate approximation of the fermionic system for this observables.
The effect of the sign on the stiffness is in fact smaller than the error-bars of our simulations, so within the accuracy of our simulations the quadrupling transition of the sign-full and sign-free models coincide.

On the left side of figure \ref{fig:sign-comp} we can see the equivalent plot of the relative difference of the smaller composite stiffness at $\beta = 4$ which for the sign-free model is below the superfluid-quadrupling transition and thus within the superfluid state.
For this observable the difference between the sign-free simulations and the true fermionic system is around $4\%$, which shows that this observable (and by extension the whole superfluid-quadrupling transition) in fact is sensitive to the fermionic sign.
This difference is large enough to significantly alter the size-dependent critical temperature and as a consequence we cannot use this model to determine the exact location of the superfluid-quadrupling transition.
However in both the sign-full and sign-free models all superfluid densities are above their critical value, showing that the state is superfluid even in the fermionic model for $\beta = 4$ and inferring that this transition occurs somewhere between $\beta = 4$ and the sign-free superfluid transition temperature.

\section{Appendix III: $T_{c2}$ for infinitely bound system}
\begin{figure}[!htb]
    \includegraphics[width=\linewidth]{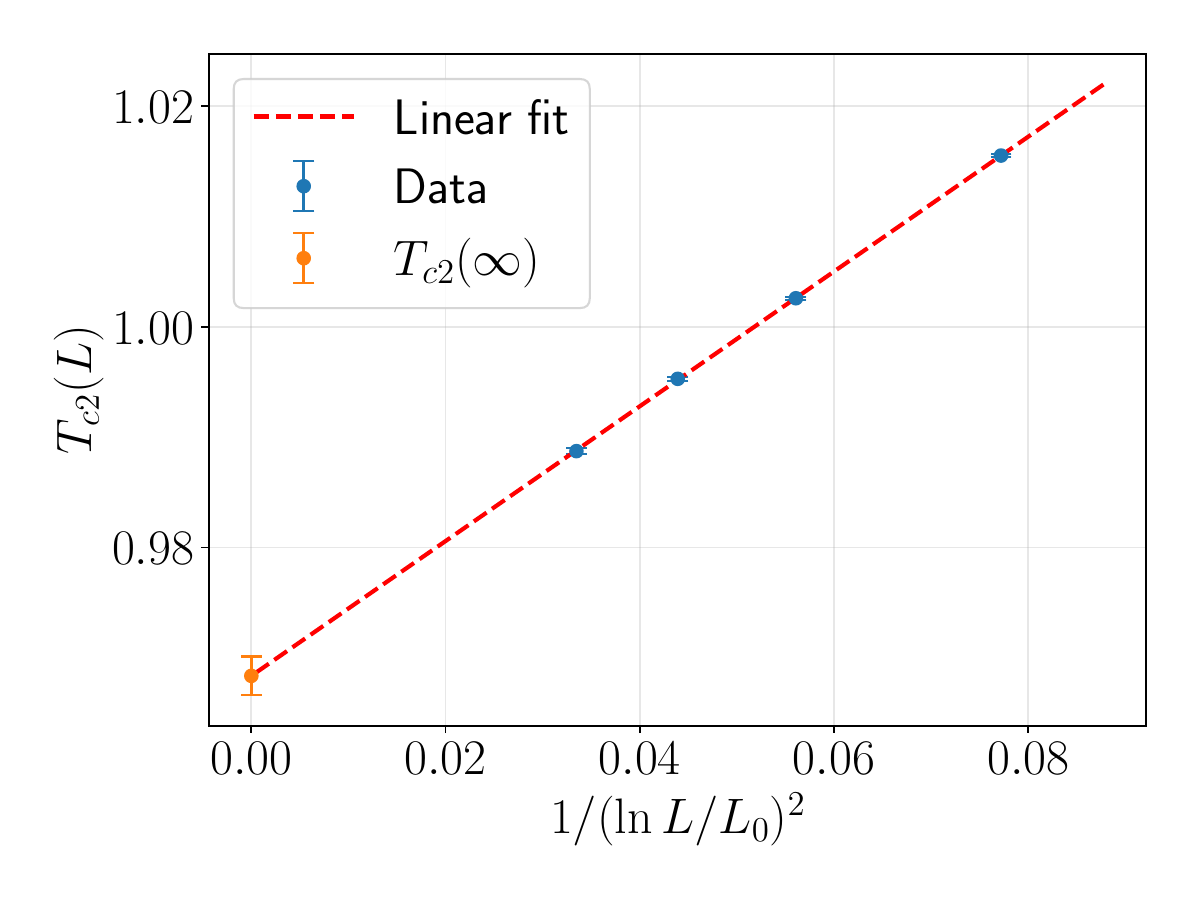}
    \caption{Critical temperature of the quadrupling-normal transition as a function of $1/(\ln L/L_0)^2$ where $L$ is the linear system size and $L_0$ is a fit parameter from the finite-size scaling 
    for a system of infinitely bound pairs ($U = -\infty$). 
    Points show simulation data with error bars and the solid lines represent the linear fit to the universal scaling form.
    This fit corresponds to the parameters $T_{BKT} \approx 0.9684 \pm 0.0017$ (equivalent to $\beta_{BKT} \approx  1.033$), $A \approx 0.6103 \pm 0.0896$ and $L_0 \approx 0.5792 \pm 0.1160$.}
    \label{fig:boson_quad_scaling}
\end{figure}

The critical temperature for the infinitely bound system is shown in Figure \ref{fig:boson_quad_scaling}, and is estimated at $T_{c2} \approx 0.97$.
This is significantly larger than the transition temperature for the moderately bound system, indicating that the motion of single particles has a large effect on the transition.

\end{document}